\documentstyle[floats,aps,twocolumn,psfig]{revtex}

\renewcommand{\bar}{\overline}
\def\bC{\begin{center}\hspace{.1mm}}
\def\eC{\vspace*{2mm}\end{center}}
\def\eps{\varepsilon}
\begin{document}
\title{\vspace*{-5mm}\hfill
{\normalsize submitted to Proc. of 8th Int. Conf. on Vibrations at
Surfaces}
        \vspace{5mm}\\
Anomalies in He atom scattering spectra\\
of the H-{}covered Mo\,(110) and W\,(110) surfaces}

\author{Bernd Kohler, Paolo Ruggerone, and Matthias Scheffler}
\address{
Fritz-Haber-Institut der Max-Planck-Gesellschaft,\\
Faradayweg 4-6, D-14\,195 Berlin-Dahlem, Germany}
\date{May 1, 1996}
\maketitle
\begin{abstract}
Helium atom scattering (HAS) studies of the H-{}covered 
Mo\,(110) and W\,(110) surfaces reveal a twofold anomaly in the 
respective dispersion curves. 
In order to explain this unusual behavior we performed  
density-{}functional theory calculations of the atomic and 
electronic structure, 
the vibrational properties, and the electronic susceptibility 
of those surfaces.
Our work provides evidence for hydrogen adsorption 
induced Fermi-{}surface nesting. 
The respective nesting vectors are in excellent agreement with
the HAS data and recent angle resolved photo\-emission experiments 
of the H-{}covered alloy system Mo$_{0.95}$Re$_{0.05}$\,(110). 
Also, we investigated the electron-{}phonon coupling and discovered 
that the Rayleigh phonon frequency is lowered for those critical 
wave vectors compared to the clean surfaces. 
Moreover, the smaller indentation in the HAS spectra can 
be clearly identified as a Kohn anomaly. 
Based on our results for the susceptibility and the recently 
improved understanding of the He scattering mechanism we argue 
that the larger anomalous dip is due to electron-{}hole excitations 
by the He scattering. 
\end{abstract}

\section{Introduction}

\begin{figure}\bC
\psfig{file=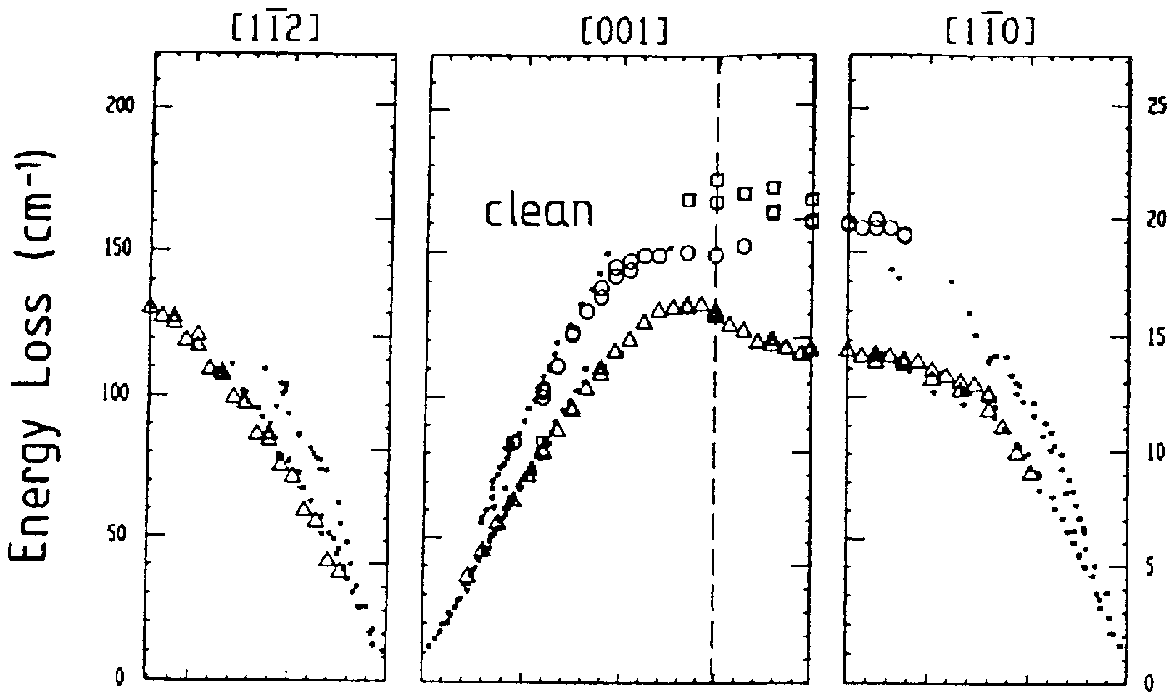,width=7.5cm}
\psfig{file=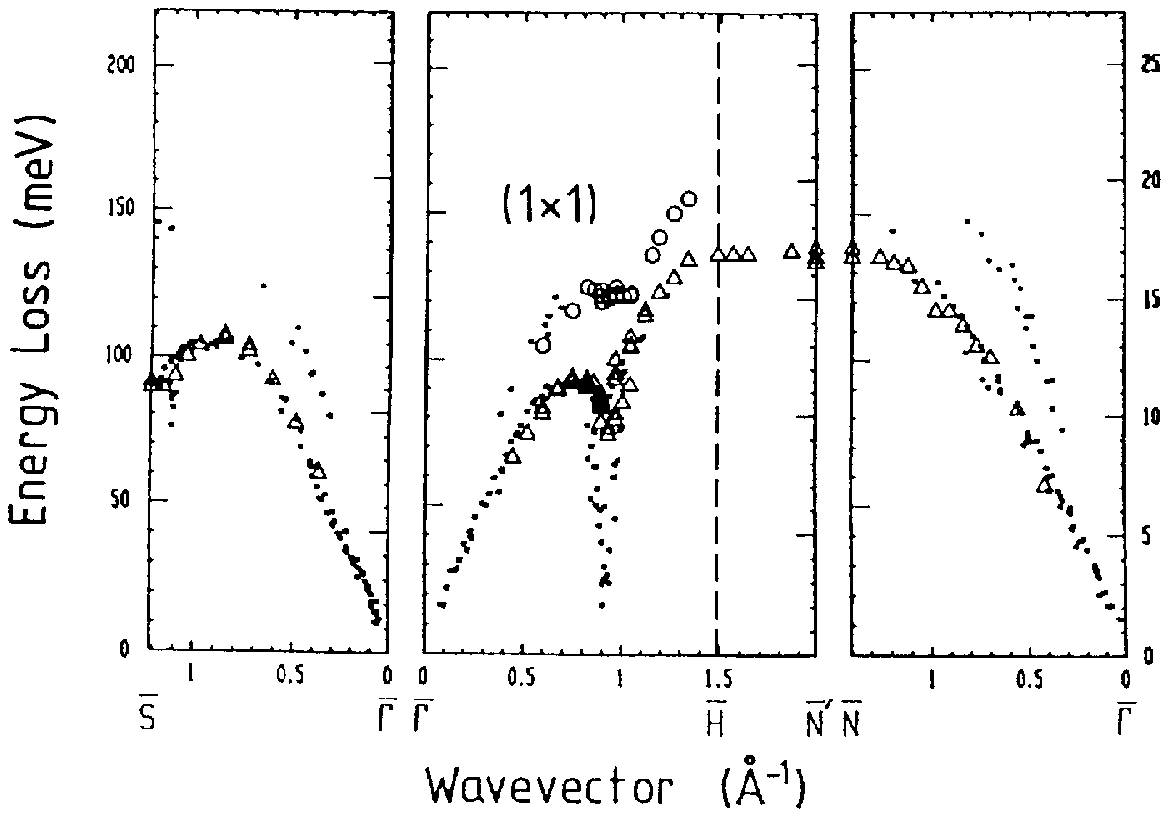,width=7.5cm}
\eC
\caption{
HREELS spectra of the clean W(110) surface (upper panel) and 
its H-{}covered $(1\times 1)$ phase (lower panel)
~\protect{\cite{bald94,bald95}}: 
Shown are the dispersion of the Rayleigh 
wave (triangles) and the longitudinally polarized surface phonons 
of the first (circles) and second (squares) layer. 
The dots represent the results of the HAS 
measurements~\protect{\cite{hulp92a,lued94}}.}
\label{Fanom}
\end{figure}
Much interest and many questions have been raised by 
the recent discovery of deep and extremely sharp indentations in the 
surface phonon spectra of H/Mo\,(110) and H/W\,(110) at full hydrogen 
coverage~\cite{hulp92a,lued94} 
(see Fig.\,\ref{Fanom}).
Those anomalies are seen at an incommensurate wavevector, 
$\bar Q^{\rm c,1}$, 
along the $[001]$ direction (${\overline{\Gamma H}}$) and
additionally 
at the commensurate wavevector 
$\bar Q^{\rm c,2} = {\bar S}$ at the boundary of the surface 
Brillouin zone (SBZ).
Out of the ordinary Rayleigh mode two simultaneous anomalies 
develop at those points. 
One, $\omega_1$, is extremely deep, and is only seen by helium atom  
scattering (HAS). 
The other, $\omega_2$, is instead soft, and is observed by both HAS 
and high resolution electron energy loss spectroscopy 
(HREELS)~\cite{bald94,bald95}. 

A pronounced but less sharp softening of surface phonons is 
also seen on the clean (001) surfaces of W~\cite{erns92a} and 
Mo~\cite{hulp89}. There, the effect can be explained in terms of a 
nested structure of the Fermi surface~\cite{smit90,chun92}. 
However, for the phonon anomalies of H/W\,(110) and H/Mo\,(110)
such a 
connection between the electronic structure and the vibrational 
properties seemed to be non-{}existent since 
angular resolved photo\-emission (ARP) experiments of the two systems 
gave no evidence for nesting vectors comparable to the HAS determined 
critical wave vectors~\cite{jeon8X,gayl89}. 
An alternative model which links the 
phonon anomalies to the motion of the hydrogen adatoms~\cite{bald94}
has
to be ruled out because the HAS spectra 
remain practically unchanged when deuterium is adsorbed instead of 
hydrogen~\cite{hulp92a,lued94}.

\begin{figure}\bC
\psfig{figure=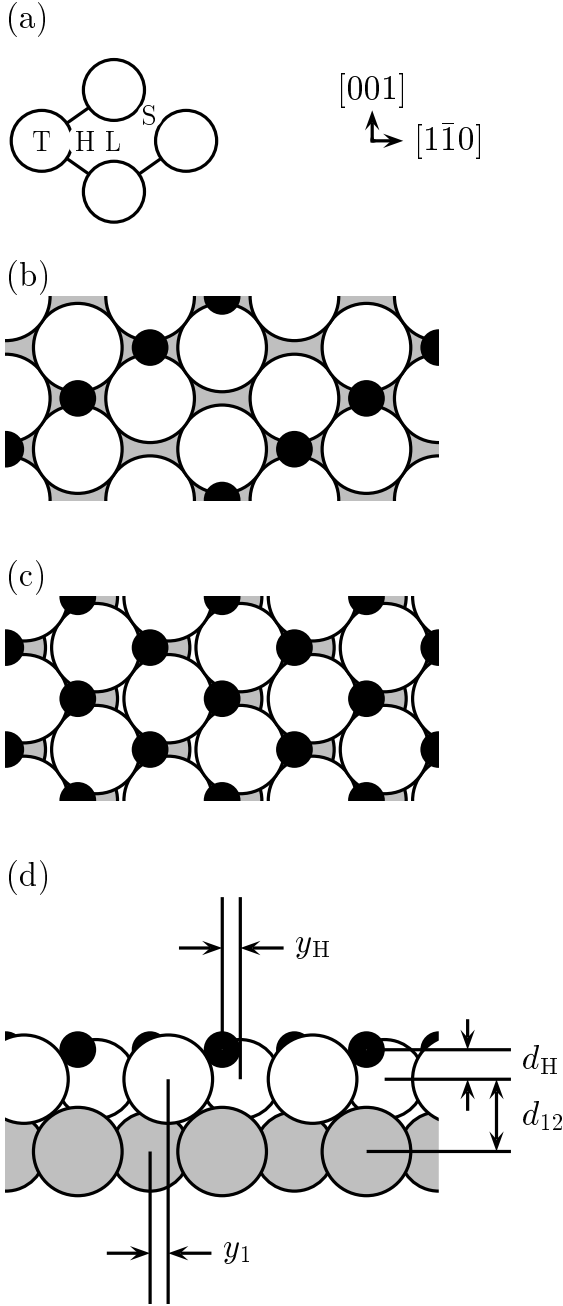,width=7.5cm}
\eC
\caption{(a)~Adsorbate positions within the 
$p(1\times 1)$ unit cell of the W\,(110) and Mo\,(110) surfaces: 
Shown are the long-{}bridge~(L), short-{}bridge~(S), hollow~(H), 
and on-{}top~(T) site. 
(b-c)~Structure of the H-{}covered W\,(110) surface as 
suggested by Chung {\em et al.} \protect{\cite{chun86}}: 
(b)~low H-{}coverage $\Theta< 0.5$\,ML; 
(c)~top-layer-shift reconstruction for higher H-{}coverages 
$\Theta>0.5$\,ML. 
The white (shaded) circles represent the W/Mo atoms of the 
surface (subsurface) layer. 
The H positions are indicated by full dots. 
(d)~Structure parameters used in Table~\protect{\ref{Tstructure}}. 
}
\label{Fadso}
\end{figure}
Another puzzle is added by the observation of a symmetry loss in 
the low-{}energy electron diffraction (LEED) 
pattern of W\,(110) upon H-{}adsorption. 
The phenomena may be caused by a H-{}induced displacement of the top 
layer W atoms along the $[\overline{1}10]$ direction~\cite{chun86} 
(see Figs.\,\ref{Fadso}b and c). 
Similar studies for H/Mo\,(110) do not provide any evidence for
a top-{}layer-{}shift reconstruction~\cite{altm87}.

Within a first-{}principles approach, 
the goal of our work is to give a comprehensive explanation for the 
observed anomalous behavior and clarify the confusing picture 
drawn by the different experimental 
findings.

\section{Method}
We perform density-{}functional theory (DFT) calculations 
using the local-{}density approximation (LDA) for the
exchange-{}correlation energy functional~\cite{cepe80}. 
For the self-{}consistent solution of the Kohn-{}Sham equations
a full-{}potential linearized augmented plane-{}wave (FP-{}LAPW) 
code~\cite{blahXX} 
is employed which we enhanced by the direct calculation of 
atomic forces~\cite{yu91,kohl96a}. 
Within a damped molecular dynamics approach this enables an 
efficient determination of fully relaxed atomic geometries and 
frozen-{}phonon energies. 
For methodical details we refer to Ref~\cite{kohl95}. 
In the case of the W surfaces the valence and semi-{}core (core)
electrons are treated scalar (fully) relativistically while
for Mo all electrons are handled non-{}relativistically.
The in-{}plane lattice constants $a_{\rm W}^{\rm theo}=3.14\,$\AA\ and
$a_{\rm Mo}^{\rm theo}=3.13\,$\AA\  are calculated without 
including zero-{}point vibrations. 
They are in good agreement with the respective measured bulk lattice
parameters  at room temperature which are 
3.163\,\AA\ and 3.148\,\AA\ for W~\cite{shah71} 
and Mo~\cite{kata79}, respectively.

We performed systematic tests comparing the LDA and the 
generalized gradient approximation~\cite{perd92} 
exchange-{}correlation functionals. 
For the quantities reported in this paper (total energy differences
of  
different adsorption sites, bond lengths, etc.) 
both treatments give practically the same results. 
Also, we checked the ${\bf k_{\|}}$-{}point convergence: 
In the case of atomic and electronic structure calculations a 
two-{}dimensional linear 
mesh of 64 ${\bf k_{\|}}$-{}points within the $(1\times 1)$ SBZ
gives stable 
results. 
For the evaluation of frozen-{}phonon energies a set 
of 56\,${\bf k_{\|}}$-{}points is employed within the SBZ of the
enlarged $(1\times 2)$ and $(2\times 1)$ surface cells. 

\section{Results}
\subsection{Atomic Structure}

\begin{table}[tb]
\caption{
Relaxation parameters for the clean and H-{}covered
(110) surfaces of Mo and W (see also Fig.\,\protect{\ref{Fadso}}d). 
For each system the results from a nine-{}layer slab calculation 
(first line) and a LEED analysis 
(second line)~\protect{\cite{arno96,arno96b}} are presented. 
The height of the hydrogen above the surface and its 
$[\overline{1}10]$ offset from the [001] bridge position
are denoted by $d_{\rm H}$ and $y_{\rm H}$, respectively. 
The shift of the surface layer with respect to the substrate
is $y_1$.
The parameters $\Delta d_{ij}$ describe the percentage change of the
interlayer distance between the $i$-{}th and the $j$-{}th substrate
layers with
respect to the bulk interlayer spacing $d_0$. 
The numerical accuracy of the theoretically obtained parameters is 
$\pm0.015\,$\AA$\approx \pm 0.7\%d_0$.}
\begin{tabular}{cccccc}
$y_{\rm H}$
&$d_{\rm H}$
&$y_1$
&$\Delta d_{12}$
&$\Delta d_{23}$
&$\Delta d_{34}$
\\
(\AA)
&(\AA)
&(\AA)
&($\%d_0$)
&($\%d_0$)
&($\%d_0$)
\\
\hline \multicolumn{6}{c}{\underline{Mo\,(110)}}\\
$-$
&$-$
&$-$
&$-5.0$
&$+0.7$
&$-0.3$
\\
$-$
&$-$
&$-$
&$-4.0\pm0.6$
&$+0.2\pm0.8$
&$0.0 \pm1.1$
\\
\hline \multicolumn{6}{c}{\underline{H/Mo\,(110)}}\\
0.62
&1.09
&0.04
&$-3.3$
&+0.3
&$-0.2$
\\
$0.55\pm 0.4$
&$1.3\pm 0.3$
&$0.0\pm 0.1$
&$-2.0\pm 0.4$
&$0.0\pm0.5$
&$0.0\pm0.8$
\\
\hline \multicolumn{6}{c}{\underline{W\,(110)}}\\
$-$
&$-$
&$-$
&$-$3.6
&$+$0.2
&$-$0.3
\\
$-$
&$-$
&$-$
&$-3.1\pm 0.6$
&$0.0\pm0.9$
&$0.0\pm1.0$
\\
\hline \multicolumn{6}{c}{\underline{H/W\,(110)}}\\
0.67
&1.09
&0.02
&$-1.4$
&$-0.3$
&$-0.1$\\
$0.56\pm0.4$
&$1.2\pm0.25$
&$0.0\pm0.1$
&$-1.7\pm0.5$	
&$0.0\pm0.6$
&0.0$\pm$0.9
\end{tabular}
\label{Tstructure}
\end{table}
The first step in our study is to determine the atomic structures 
of the clean and hydrogen-{}covered (110) surfaces. 
The relaxation parameters presented in 
Table~\ref{Tstructure} are calculated for a nine-{}layer slab 
(the parameters are defined in Fig.~\ref{Fadso}d). 
The results, which are remarkably similar for Mo and W, are in
excellent 
quantitative agreement with a recent LEED
analysis~\cite{arno96,arno96b}. 

\begin{figure}\bC
\psfig{figure=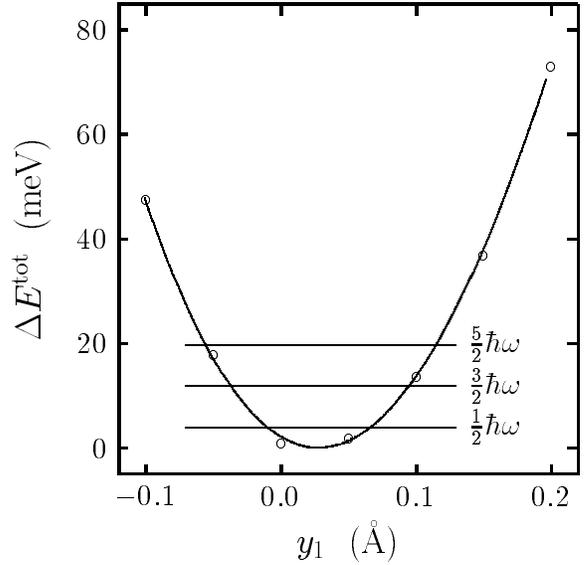,width=7.5cm}
\eC
\caption{Change of total energy $E^{\rm tot}$ versus 
top-{}layer-{}shift $y_1$: 
For each data point the whole surface is relaxed keeping only 
the substrate $[1\bar10]$-{}coordinates $y_1$ and $y_2=y_3=0\,$\AA~ 
fixed.
Also shown are the first three oscillator eigenstates calculated 
from a harmonic expansion of the total energy $E^{\rm tot}(y_1)$.}
\label{Fshift}
\end{figure}
As the energetically most favorable hydrogen adsorption site on both 
W\,(110) and Mo\,(110) we identify a quasi-{}threefold position 
(indicated as H in Fig.\,\ref{Fadso}a). 
Our investigations also throw light on the suggested 
model of a H-{}induced structural change: 
For both materials the calculated shift $y_1$ is only of the order of 
0.01\,\AA\ and thus there is no evidence for a pronounced 
top-{}layer-{}shift reconstruction. 
Moreover, this subtle change in the surface geometry is unlikely to 
be resolved experimentally due to zero-{}point vibrations. 
This aspect becomes evident by evaluating the total energy 
with respect to a rigid top-{}layer-{}shift. 
In Fig.~\ref{Fshift} we present the results of such a calculation 
for H/W\,(110) and depict the first three vibrational eigenstates 
obtained from a harmonic expansion of the total energy
$E^{\rm tot}(y_1)$; 
for H/Mo\,(110) the energetics is similar. 
Since we have $k_{\rm B}T\approx 25$\,meV at room temperature 
thermal fluctuations of $y_1$ are of the order of 0.1\,\AA. 
This is considerably larger than the theoretically predicted 
absolute value of $y_1$. 

\subsection{Electronic Structure} 

\begin{figure}\bC
\psfig{file=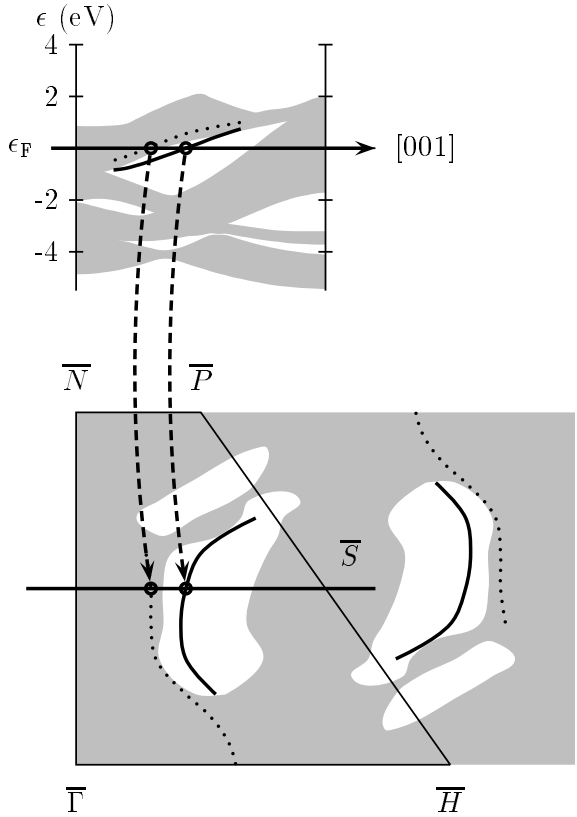,width=7.5cm}
\eC
\caption{Band structure (upper figure) and Fermi surface (lower
figure) 
of clean Mo\,(110) (dotted curves) and H/Mo\,(110) (solid curves). 
The shaded areas represent the projection of the bulk Fermi
surface onto the 
SBZ of the (110) surface.}
\label{Fband}
\end{figure}
The similarities between Mo\,(110) and W\,(110) found for the 
surface geometries continue when we compare the electronic 
structures. 
For both systems the H adsorption alters the surface potential and 
induces the shift of a band with $(d_{3z^2-r^2},d_{xz})$ 
character to higher binding energies~\cite{kohl95,rugg95}. 
Due to this effect which is illustrated in Fig.~\ref{Fband} the 
Fermi line associated with this band is moved into the band 
gap of the surface projected band structure, and the respective 
states become true surface states. 

\begin{figure}\bC
\psfig{file=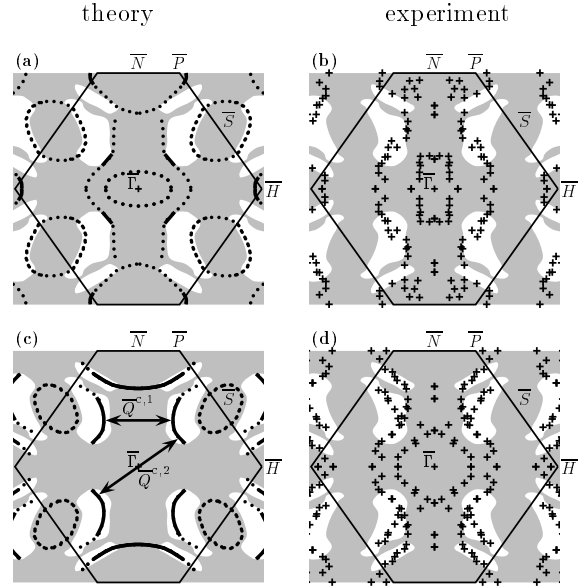,width=7.5cm}
\eC
\caption{
Theoretical Fermi surfaces of the (a) clean 
and (c) H-{}covered Mo\,(110) surface. 
The solid (dotted) lines denote surface resonances or surface 
states which are localized by more than 60 \% (30 \%) in the MTs 
of the two top Mo layers. 
Shaded areas represent the (110) projected theoretical Mo bulk 
Fermi surface. 
The arrows ${\bar Q}^{\rm c,1}$ and ${\bar Q}^{\rm c,2}$ 
are the critical wave vectors predicted by the calculations. 
Also presented are data points (+) 
which stem from an ARP study of the (b) clean and (d) H-{}covered 
Mo$_{0.95}$Re$_{0.05}$\,(110) surface~\protect{\cite{okad96}}. }
\label{Ffermi}
\end{figure}
The shifted $(d_{3z^2-r^2},d_{xz})$ band is characterized by a high 
density of states at the Fermi level. 
More important, the new Fermi contour shows pronounced nesting 
features (see Fig.~\ref{Ffermi}c). 
The magnitudes of the calculated nesting vectors and HAS measured 
critical wavevectors along $\overline{\Gamma S}$ and 
$\overline{\Gamma H}$ are listed in Table~\ref{Tnesting}. 
For both systems the agreement between theory and experiment is 
excellent.
\begin{table}[b]
\caption{Theoretical Fermi-{}surface nesting vectors compared to
critical 
wavevectors obtained by HAS and HREELS 
ex\-periments~\protect{\cite{hulp92a,lued94,bald94}}.
 }
\begin{tabular}{llll}
direction       &system         &\multicolumn{2}{c}{
$|\bar Q^{\rm c}|\ \
(\mbox{\AA}^{-1})$}\\
                                \cline{3-4}
                &               &theory &experiment\\
\hline  $\overline{\Gamma H}\ \ $
                        &H/Mo\,(110)      &0.86   &0.90 \\
                        &H/W\,(110)       &0.96   &0.95 \\
\hline  $\overline{\Gamma S}\ \ $
                        &H/Mo\,(110)      &1.23   &1.22\\
                        &H/W\,(110)       &1.22   &1.22\\
\end{tabular}
\label{Tnesting}
\end{table}

Our results are at variance with the photoemission studies by 
Kevan and coworkers~\cite{jeon8X,gayl89} (see Fig.~2 in 
Ref.~\cite{kohl95}). 
However, in view of more recent ARP measurements~\cite{okad96} 
which we compare to our results in Fig.~\ref{Ffermi} 
we dare to suggest that those differences may be due to 
problems in the experimental analysis. 
The later work deals with the (110) surface 
of the alloy Mo$_{0.95}$Re$_{0.05}$ but the surface physics of 
both systems Mo$_{0.95}$Re$_{0.05}$\,(110) and Mo\,(110) 
should be practically the same 
because in both cases the top layer consists 
only of Mo atoms. 
This assumption is also backed by test calculations where we 
simulated a Mo$_{0.95}$Tc$_{0.05}$ alloy be the virtual 
crystal approximation and found only 
minor quantitative changes. 

From Fig.~\ref{Ffermi} it becomes clear that in particular for 
the important $(d_{3z^2-r^2},d_{xz})$ surface band, which was 
not seen by Kevan's group, experiment and DFT now agree very well. 
There are however differences: 
Theory predicts bands centered at $\bar S$ which are not seen by 
ARP whereas the band circle at $\bar \Gamma$ in Fig.~\ref{Ffermi}d 
is only observed experimentally. 
Also, in the calculations we find an elliptical band centered at the 
$\bar \Gamma$ point whereas in ARP the same band has the shape 
of a rectangle. 
Those discrepancies are probably due to the fact that it is rather 
difficult in theory (and experiment!) to clearly distinguish 
between surface states and surface resonances.  
Additionally, due to interactions between the two surfaces within a 
slab system the $\bf k_{\|}$-{}space location of a 
calculated surface resonance which is not strongly localized 
at the surface cannot be determined accurately.  
Furthermore, we note that DFT is not expected to describe the 
Fermi surface accurately. 
Nevertheless, the important theoretical finding of a H-{}induced 
surface state which can be related to the HAS anomalies is now 
verified by experiment. 

\subsection{Vibrational Properties}
At this point, the following question arises: 
How does the surface react to the apparent electronic
instability due to the strong nesting features? 
In the case of nesting the surface phonons take the burden to furnish 
energy and momentum  to the scattering process of electrons at the 
Fermi level. 
This leads to a softening of the related phonons. 
If the electron-{}phonon coupling is strong and the energetic cost 
of a surface distortion is small this softening could even trigger 
a reconstruction combined with the build-{}up of a charge-{}density 
wave as in the case of the (001) surfaces of W~\cite{erns92a} and 
Mo~\cite{hulp89}. 
It is clear that one needs to perform frozen-{}phonon calculations in 
order to determine the actual strength of the coupling and the 
induced phonon softening~\cite{kohl96b}.  

\begin{table}
\caption{Comparison of calculated frozen-{}phonon energies and
experimental values obtained by 
HAS~\protect{\cite{hulp92a,lued94}} and 
HREELS~\protect{\cite{bald94}}.
The theoretical phonon energies are obtained using a five-{}layer
slab.  
Their numerical accuracy is about $\pm 1$\,meV.}
\begin{tabular}{llll}
phonon          &system         &\multicolumn{2}{c}{$E^{\rm ph}$\ \
(meV)}\\
                                \cline{3-4}
                &               &theory &experiment\\
\hline
$\bar N$       &W\,(110)         &15.4   &14.5\\
                &H/W\,(110)       &17.6   &17.0\\
\hline
$\overline{S}$       &Mo\,(110)        &22.7   &$\sim$21\\
                &H/Mo\,(110)      &17.2   &$<$16\\
                \cline{2-4}
                &W\,(110)         &18.3   &16.1\\
                &H/W\,(110)       &12.0   &11.0\\
\end{tabular}
\label{Tphonon}
\end{table}
Fortunately, at $\bar Q^{\rm c,2}=\bar S$ such a calculation is 
particularly convenient, since at the zone boundary point $\bar S$ 
the Rayleigh-{}wave polarization is purely vertical, and the second
layer 
is immobile by symmetry.
The calculated frequencies for the $\bar S$-{}point and a second 
zone boundary point, $\bar N$, are presented in 
Table~\ref{Tphonon}.
For the latter our results reproduce the experimentally
observed increase of the Rayleigh wave frequency as hydrogen is
adsorbed. 
At the ${\bar S}$ point we find that those stiffening effects are 
over-{}compensated: 
The strong coupling to electronic states at the Fermi level leads 
to a lowering of the phonon energy in good agreement 
with the experimental results. 

\begin{figure}\bC
\psfig{file=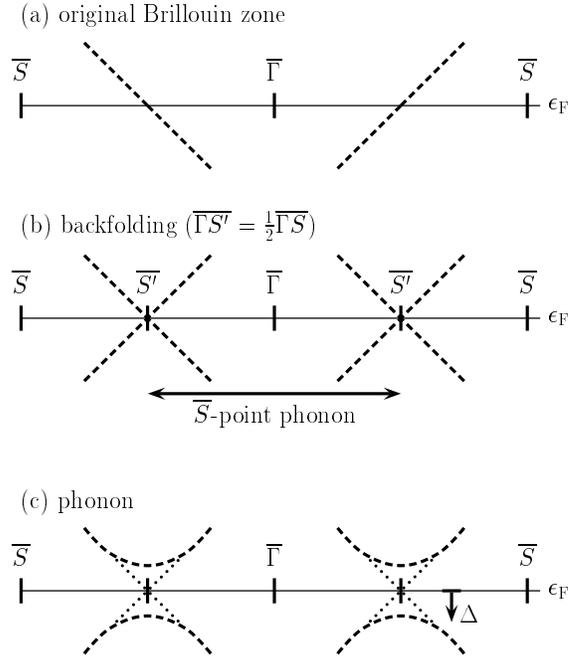,width=7.5cm}
\eC
\caption{Schematic representation of the mechanism 
which causes the Kohn anomaly on H/Mo\,(110) and H/W\,(110). 
Shown is the bandstructure along $\overline{S\Gamma S}$ parallel 
to the nesting vector $\bar Q^{\rm c,2}$ in 
Figs.\,\protect{\ref{Ffermi}}. 
The form of the \protect{($d_{3z^2-r^2},d_{xz}$)} band is 
indicated by dashed lines.}
\label{Fkohn}
\end{figure}
In Fig.~\ref{Fkohn} we schematically illustrate the mechanism 
which is responsible for this effect. 
It is, in fact, a text-{}book example of a Kohn 
anomaly due to Fermi-{}surface nesting~\cite{migd58,kohn59}: 
(a) Within the unperturbated system the ($d_{3z^2-r^2},d_{xz}$) band 
cuts the Fermi level exactly midway between $\bar \Gamma$ and
$\bar S$. 
(b) The $\bar S$-{}point phonon couples to the states at the Fermi
level 
and causes the backfolding of the SBZ. 
(c) 
The nuclear distortion associated with the $\bar S$-{}point phonon 
modifies the surface potential and hence removes the degeneracy at 
the new zone boundary point $\bar{S'}$. 
The occupied states are shifted to lower energies. 
This amounts to a negative contribution of the electronic band 
structure energy to the total energy and thus a lowering of the 
phonon energy. 

A frozen-{}phonon study for the second nesting vector 
along $\overline{\Gamma H}$ is not performed because 
$\bar Q^{\rm c,1}$ is highly non-{}commensurate; 
a calculation would be very expensive.  
However, since the character of the $(d_{3z^2-r^2},d_{xz})$ band 
does not change when shifting from the $\overline{\Gamma S}$ to 
the $\overline{\Gamma H}$ nesting we expect similar results. 

The calculation of the electron-{}phonon interaction
at the Mo\,(110) and W\,(110) surfaces and its change 
due to hydrogen adsorption pinpoints the phonon character of the 
small anomaly $\omega_2$ and  identifies  the interplay between 
the electronic structure and the vibrational spectra of the 
transition metal surfaces. 
Thus, our results which were recently confirmed by 
Bungaro~\cite{bung96} within the framework of perturbation DFT 
clearly support the interpretation that the small dip 
observed by both HAS and HREELS is due to a Kohn anomaly. 

\subsection{Electron-{}hole excitations}
For the deep and sharp anomaly $\omega_1$ the interpretation  
is less straightforward, particularly since it is only seen 
in the HAS spectra. 
One needs to understand the nature of rare-{}gas scattering 
in order to explain this unusual behavior. 

In a recent study we found that those scattering processes 
are significantly more complicated (and more interesting)
than hitherto assumed~\cite{pete96}: 
The reflection of the He atom happens in front of the surface
at a distance of 2--3\,\AA. 
More important, it is not the {\em total} electron density of the 
substrate surface which determines the interaction but the electronic 
wave functions close to the Fermi level.  
In the case of the H/W\,(110) and H/Mo\,(110) systems 
it is thus plausible to assume that the He atom couples directly 
to the $(d_{3z^2-r^2},d_{xz})$ surface states mentioned above and 
excites electron-{}hole pairs. 

In order to analyze the spectrum of those excitations in some more 
detail we evaluate the imaginary part of the electronic 
susceptibility $\chi({\bf q_{\|}},\hbar\omega)$ within 
random phase approximation (RPA):
\begin{eqnarray}
\nonumber
\lefteqn{\mbox{Im}\ \chi({\bf q_{\|}},\hbar\omega) = }\\ \nonumber
&= \mbox{Im}&\frac{A}{4\pi^2}
 \int_{\rm SBZ} d{\bf k_{\|}}\
w_{\bf k_{\|}+\bf q_{\|}} w_{\bf k_{\|}}\ 
(f_{\bf k_{\|}+ q_{\|}}-f_{\bf k_{\|}})\times \\ \nonumber
&&\hspace{2cm} \times \frac
{\displaystyle 1}
{\displaystyle
\eps_{\bf k_{\|}+q_{\|}}-\eps_{\bf k_{\|}}-\hbar\omega -i\eta}
\\ 
&=& \frac{A}{4\pi^2}
\int_{\rm SBZ} d{\bf k_{\|}}\
w_{\bf k_{\|}+\bf q_{\|}} w_{\bf k_{\|}}\
(f_{\bf k_{\|}+\bf q_{\|}}-f_{\bf k_{\|}})\times \\ \nonumber
&&\hspace{2cm} \times \delta\left( \eps_{\bf k_{\|}+\bf q_{\|}}
-\eps_{\bf k_{\|}}-\hbar\omega\right)
\end{eqnarray}
with $\frac{A}{4\pi^2}$ being the ${\bf k_{\|}}$-{}space density. 
In our approach we use all eigenvalues $\epsilon_{\bf k_{\|}}$ and 
occupation numbers $f_{\bf k_{\|}}$ obtained via a nine-{}layer 
slab calculations. 
Also, we take into account the localization $w_{\bf k_{\|}}$ of
the respective 
state at a distance of 2.5\,\AA\ in front of the substrate 
surface. 

\begin{figure}\bC
\psfig{file=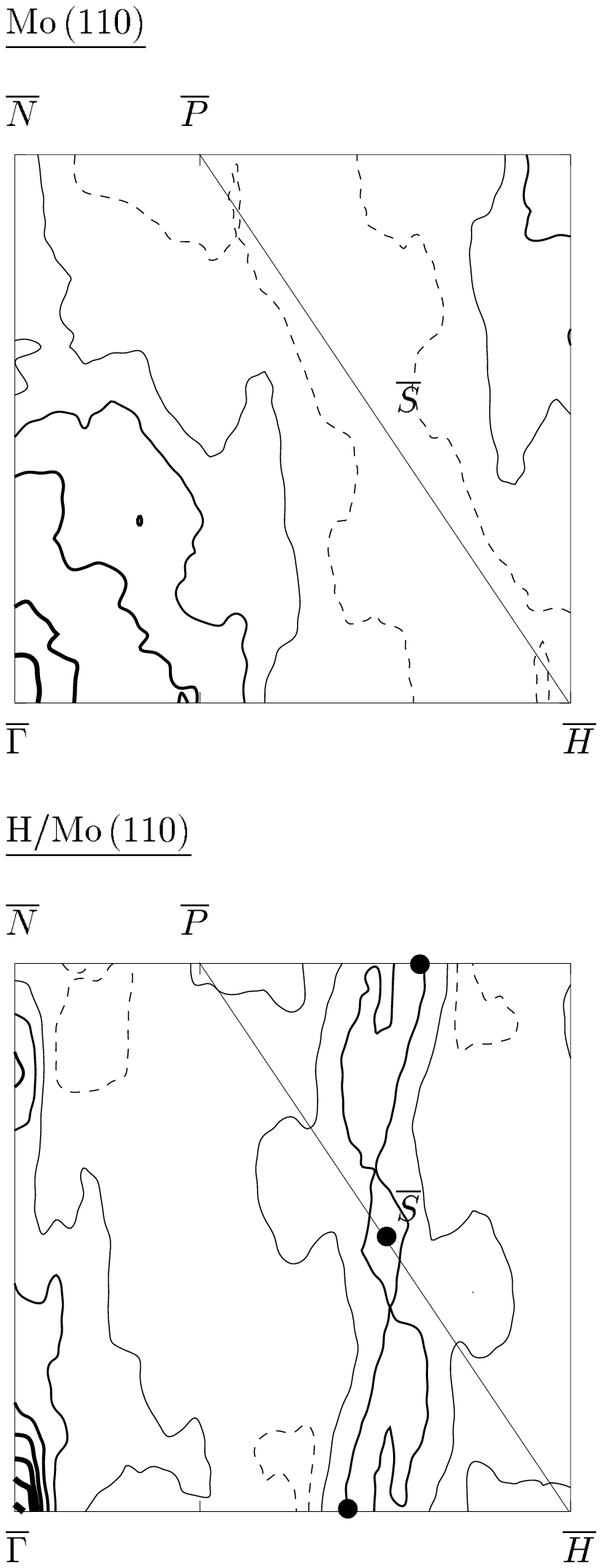,width=7cm}
\eC
\caption{Imaginary part of the electronic RPA susceptibility 
$\chi({\bf q_{\|}},\hbar \omega)$ of Mo\,(110) and H/Mo\,(110) 
calculated for $\hbar \omega = 0.5$\,mRy $\approx 7$\,meV. 
The dashed line represents a value of 3\,a.u.\ while each full 
line denotes an additional increase by 1\,a.u. (with increasing
line width).
The positions of the HAS measured anomalies within 
the SBZ are indicated by dots.}
\label{FSusc}
\end{figure}
The results obtained for Mo\,(110) and H/Mo\,(110) are presented in 
Fig.~\ref{FSusc}: 
For the clean surface we find that the intensity of 
the electron-{}hole excitations decreases continuously 
as we move away from the $\bar \Gamma$ point towards the zone 
boundaries. 
This relatively smooth behavior of the susceptibility 
is modified considerably when H is adsorbed on the clean 
surface: 
Pronounced peaks appear, in particular one which stretches from 
the $\overline{\Gamma H}$ line through the SBZ to the symmetry point 
$\bar S$. 
This is in excellent agreement with the HAS results. 
Also, there is a peak located at the second half of the 
$\overline{\Gamma N}$ symmetry line.
It cannot be 
associated with an anomaly in the HAS spectrum. 
This result may be due to the fact that selection 
rules for the interaction between the scattering He atoms and 
the electron-{}hole excitations are not considered. 

In view of the simplicity of the approach, 
the calculated spectrum of electron-{}hole excitations however 
seems to be well-{}connected to the HAS measurements. 
Moreover, it is surprising how detailed the Fermi surface nesting 
is reflected in the susceptibility at such a large distance from 
the surface where the HAS scattering process takes place. 
This finding supports the earlier suggestion~\cite{kohl95}
that the giant indentations, seen only in HAS, 
are due to the excitation of electron-{}hole pairs.
By contrast, in HREELS experiments the electrons are directly 
scattered at the atoms and thus only measure the phonon spectrum. 
Therefore, the strong anomaly is invisible for HREELS. 

\section{Summary}
In conclusion, the major results of our work are as follows: 
We demonstrate that both the W\,(110) and the Mo\,(110) 
surface do not show a pronounced top-{}layer-{}shift reconstruction
upon adsorption of hydrogen,
a result confirmed by a recent LEED analysis~\cite{arno96}. 
Also, we give a consistent explanation for the H-{}induced 
anomalies in the HAS spectra of W\,(110) and Mo\,(110). 
The H adsorption induces surface states of $(d_{3z^2-r^2},d_{xz})$ 
character that show pronounced Fermi-{}surface nesting. 
The soft anomalous dip is clearly identified as a Kohn 
anomaly due to those nesting features~\cite{kohl95,rugg95,kohl96b}. 
For an understanding of the huge dip one needs to take into 
account that the scattering of rare-{}gas atoms is crucially
influenced 
by interactions with substrate surface wave functions at the Fermi 
level~\cite{pete96}. 
The He atom couples efficiently to the H-{}induced surface 
states on H/W\,(110) and H/Mo\,(110). 
We therefore conclude that the deep branch of the anomaly   
is predominantly caused by a direct excitation of
electron-{}hole pairs during the scattering process. 

We hope that our first-{}principles study stimulates some more 
experimental work: 
For instance, the liquid-{}like behavior of the H adatoms 
observed in the HREELS measurements~\cite{bald94} is still not 
understood. 
One would also like to know whether the HAS and HREELS spectra of 
the Mo$_{0.95}$Re$_{0.05}$\,(110) alloy system reveal H-{}induced 
anomalies as expected.  
Finally, we call for a new ARP study of the clean and H-{}covered 
Mo\,(110) and W\,(110) surface and think that scattering 
experiments with atoms or molecules like Ne or H$_2$ might provide 
additional insight into the interesting behavior of these 
surfaces. 

We thank Erio Tosatti for stimulating discussions. 

\renewcommand{\,}{ }

\end{document}